\documentclass[12pt,twocolumn]{iopart}

\expandafter\let\csname equation*\endcsname\relax
\usepackage{multicol}
\expandafter\let\csname endequation*\endcsname\relax

\usepackage{amsmath}
\usepackage{amsfonts}
\usepackage{iopams}
\usepackage{cite}
\usepackage[colorlinks=true, citecolor=blue]{hyperref}
\usepackage{graphicx}
\usepackage{hyperref}
\usepackage{graphics}
\usepackage{amssymb}
\usepackage{color}
\usepackage{subfigure}
\usepackage{bm}
\usepackage{float}
\usepackage[utf8x]{inputenc}

\begin{document}

\title[Entropic uncertainty relation in Garfinkle-Horowitz-Strominger dilation black hole]{Entropic uncertainty relation in Garfinkle-Horowitz-Strominger dilation black hole}

\author{Fariba Shahbazi$^{1}$, Soroush Haseli $^{2}$ \footnote{
Corresponding author}, Hazhir Dolatkhah $^{1}$  and Shahriar Salimi $^{1}$  }

\address{
$^{1}$ Department of Physics, University of Kurdistan, P.O.Box 66177-15175 , Sanandaj, Iran\\
$^{2}$ Faculty of Physics, Urmia University of Technology, Urmia, Iran

}

\ead{soroush.haseli@uut.ac.ir}

\vspace{10pt}
\begin{indented}
\item[]
\end{indented}

\begin{abstract}
Heisenberg's uncertainty principle is a fundamental element in quantum mechanics. It sets a bound on our ability to predict the measurement outcomes of two incompatible observables simultaneously. In quantum information theory, the uncertainty principle can be expressed using entropic measures. The entropic uncertainty relation can be improved by considering an additional particle as a memory particle. The presence of quantum correlation between the memory particle and the measured particle reduces the uncertainty. In a curved space-time, the presence of the Hawking radiation can reduce quantum correlation. Therefore, concerning the relationship between the quantum correlation and entropic uncertainty lower bound, we expect that the Hawking radiation increases the entropic uncertainty lower bound. In this work, we investigate the entropic uncertainty relation in Garfinkle-Horowitz-Strominger (GHS) dilation black hole. We consider a model in which the memory particle is located near the event horizon outside the black hole, while the measured particle is free falling. To study the proposed model, we will consider examples with Dirac fields. We also explore the effect of the Hawking radiation on the quantum secret key rate.
\end{abstract}

%

%
%
%
%

\section{Introduction}\label{sec:intro}
The uncertainty principle is one of the distinguishing features between quantum and classical theory. The Heisenberg uncertainty principle states that it is not possible to measure the location and momentum of a particle accurately \cite{Heisenberg}. Accurate measurement of one observable reduces the accuracy of another observable measurement. So far, this principle has been expressed in various ways. One of the most fundamental forms of expressing the uncertainty  principle was provided by Schrodinger \cite{Schrodinger} and Robertson \cite{Robertson}. They showed that for any arbitrary pairs of noncommuting observables $Q$ and $R$ the following relation is established for the uncertainty principle
\begin{equation}\label{Robertson}
\Delta Q\Delta R\geq \frac{1}{2}|\langle \left[Q, R\right]\rangle |,
\end{equation}
where $\Delta X=\sqrt{\langle X^{2} \rangle- \langle X \rangle^{2}}$ with $X \in \lbrace Q,R \rbrace$ is the standard deviation of  the associated observable $X$, $\langle X \rangle $ shows the
expectation value of operator $X$ and $\left[Q, R \right]=QR-RQ $.  The lower bound in Eq.(\ref{Robertson}) is state dependent which leads to a trivial bound if $|\langle \left[Q, R\right]\rangle |=0$. In quantum information theory, it has been shown that the most appropriate quantity to show  the uncertainty is  entropy. Uncertainty relations that are defined in terms of entropy are called entropic uncertainty relation (EUR). The first EUR was speculated by Deutsch \cite{Deutsch}, then improved by Kraus \cite{Kraus}, and finally proved by Maassen and Uffink \cite{Maassen}. They showed that for any arbitrary pairs of observables $Q$ and $R$ EUR can be written as 
\begin{equation}\label{Rob}
H(Q) + H(R) \geq \log_2 \frac{1}{c}
\end{equation}
where $H(Q)=\sum_i p_i \log_2 p_i$ and $H(R)=\sum_j m_j \log_2 m_j$ are the Shannon entropy, $p_i=\langle q_i|\rho|q_i \rangle$, $m_j=\langle r_j|\rho|r_j \rangle$ and $c=\max_{i,j}\lbrace |\langle q_i | r_j \rangle |^{2} \rbrace$ where $|q_i\rangle$ and $|r_j\rangle$ are  eigenstates of observables $Q$ and $R$, respectively. This statement of uncertainty principle can be described by an interesting  game between Alice and Bob. At the beginning of the game, Bob prepares the particle in a quantum state $\rho$ and sends it to Alice. In the second step, Alice and Bob agree on the measurement of two observables $Q$ and $R$ by Alice on the particle. Then Alice measures one of the two observable $Q$ or $R$ on her state and sends her  measurement choice to Bob via a classical communication channel. If Bob guesses Alice's measurement correctly, he will win the game. Bob's uncertainty about
Alice's measurement outcomes is bounded by Eq.(\ref{Rob}). In this statement of the uncertainty principle, there was only one particle. But when Bob prepares  a correlated bipartite state $\rho_{AB}$ for a two particle quantum system and sends one of the particles to Alice and keeps the other part as a quantum memory , he can guess the result of Alice's measurement more accurately. Based on this uncertainty game, Berta et
al. have presented  the EUR  in the presence of quantum memory (EUR-QM) as \cite{Berta}
\begin{equation}\label{berta}
S(Q|B)+S(R|B)\geq \log_2 \frac{1}{c}+S(A|B),
\end{equation}
where $S(Q|B)=S(\rho^{QB})-S(\rho^{B})$ and $S(R|B)=S(\rho^{RB})-S(\rho^{B})$ are the conditional von-Neumann entropies of the post measurement states
\begin{equation}
\begin{array}{l}
\rho^{QB}=\sum_i (\vert q_i \rangle \langle q_i \vert \otimes I)\rho^{AB}(\vert q_i \rangle \langle q_i \vert \otimes I), \\
\rho^{RB}=\sum_j (\vert r_j \rangle \langle r_j \vert \otimes I)\rho^{AB}(\vert r_j \rangle \langle r_j \vert \otimes I),
\end{array}
\end{equation}
and $S(A|B)=S(\rho^{AB})-S(\rho^{B})$ is the conditional von
Neumann entropy. Let's take a look at some special cases: At first, If particles $A$ and $B$ are entangled, the conditional von-Neumann entropy is negative, and Bob can guess the result of Alice's measurement with better accuracy. Second, If Bob prepares the maximally entangled state in uncertainty game then Bob can guess the result of Alice's measurement perfectly \cite{Berta}. Third, If there is no memory particle, then from Eq.(\ref{berta}), the EUR is obtained as 
\begin{equation}
H(Q)+H(R)\geq \log_2 \frac{1}{c}+S(A).
\end{equation}
Due to the presence of an additional term $S(A)$, the above EUR is tighter than Maassen and Uffink uncertainty relation. So far, several works have been done to improve the EUR \cite{8,9,10,11,12,13,14,15,16,17,18,19,20,21,22,23,24,25,26,27,28,29,30,31,32,33,34,35,36,37,38,39,40,41,42,43,44,45,46,47,48,49,50,51,52,53,54,55,56,57,58,59,60,61,62,63,64,65,66}.
In Ref. \cite{62}, the authors introduced a new bound for the EUR-QM. They showed that Bob's uncertainty about the results of Alice's measurement is bounded by 
\begin{equation}\begin{aligned}
S(Q | B)+S(R | B) &=\\
&=H(Q)-I(Q ; B)+H(R)-I(R ; B) \\
& \geq \log _{2} \frac{1}{c}+S(A)-[I(Q ; B)+I(R ; B)] \\
&=\log _{2} \frac{1}{c}+S(A | B)+\\
&+\{I(A ; B)-[I(Q ; B)+I(R ; B)]\}.
\end{aligned}\end{equation}
Based on their results the EUR-QM can be written as
\begin{equation}\label{Adabi}
S(Q | B)+S(R | B) \geq \log _{2} \frac{1}{c}+S(A | B)+\max \{0, \delta\}
\end{equation} 
where 
\begin{equation}
\delta=I(A ; B)-(I(Q ; B)+I(R ; B))
\end{equation}
and 
\begin{equation}\label{Holevo}
I(X;B)=S(\rho^{B})-\sum_x p_x S(\rho_x^{B}), \quad X \in \lbrace Q, R\rbrace,
\end{equation}
Eq.\ref{Holevo} is known as Holevo quantity, $p_x=tr_{AB}(\Pi_{x}^{A}\rho^{AB}\Pi_x^{A})$ is the probability of x-th outcome and $\rho_x^{B}=\frac{tr_{A}(\Pi_{x}^{A}\rho^{AB}\Pi_x^{A})}{p_x}$  is Bob's state after the measurement of $X$ by Alice. It is worth noting that the EUR is tighter than other EURs in the presence of quantum memory.

The EUR has a variety of applications in quantum information theory such as entanglement detection \cite{67,68,69,70}, and quantum cryptography \cite{71,72}. The security of quantum key distribution protocols can be verified using the EURs \cite{73,74}. It has been shown that the bound of EUR-QM is directly related to the quantum secret key (QSK) rate \cite{66,75}. In Ref. \cite{66}, the authors have shown that the amount of key that can be extracted by Alice
and Bob $K$ is lower bounded as
\begin{equation}
K \geq \log_2 \frac{1}{c}+\max\lbrace 0,\delta \rbrace-S(R|B)-S(Q|B),
\end{equation}
The study of the EUR-QM from a relativistic point of view has been the subject of some recent works \cite{76,77,78,79}. What is clear is that the entropic uncertainty bound decreases with increasing quantum correlation between the measured particle $A$ and the quantum memory $B$. In Refs.\cite{80,81,82,83,84}, the authors have shown that the quantum correlation decreases under the influence of Hawking radiation and so the entropic uncertainty increases. In order to study the effects of Hawking radiation on entropic uncertainty bound, we consider the most straightforward  black hole: Garfinkle-Horowitz-Strominger (GHS) dilation black hole. We also consider the Dirac fields states as examples. In this situation, a quantum state is a combination of vacuum state and excited states of Dirac fields. In GHS dilation black hole space-time, we will consider a model in which the memory particle $B$ is located somewhere near the event horizon outside the black hole, and the measured particle $A$ is free falling. When the memory particle gets closer to the event horizon, the entanglement between particle memory $B$ and measured particle $A$ decreases due to the Unruh effect. In such a situation, the lower bound of EUR-QM  increases with decreasing entanglement.  This article is organized as follows. In Sec. \ref{Sec2}, the Quantum channel interpretation of the vacuum structure for Dirac fields in the GHS dilation black hole will be reviewed. In Sec. \ref{sec3} the entropic uncertainty lower bound  (EULB) in Garfinkle-Horowitz-Strominger dilation black hole is investigated by considering some examples. Finally, conclusions are presented in Sec.\ref{sec4}.

\section{Quantum channel interpretation of the vacuum structure for Dirac
fields in the GHS dilation black hole}\label{Sec2}
In Refs.\cite{Gibbons,Hawking}, the thermal Fermi-Dirac
distribution of particles with the Hawking temperature $T=\frac{1}{8\pi(M-D)}$ has been investigated in  GHS dilation black hole \cite{Damoar}. The existence of such radiation has been described as the Hawking effect. The cosmological
parameters $M$ and $D$ represent the mass of the black hole and dilation field, respectively. Here, according to the Dirac vacuum field in
GHS dilation black hole, the global coordinates ($t,r,\theta,\phi$) is used to represent the spherically symmetric line element of the GHS  dilation black hole \cite{Wang,Garfinkle,Gareia}
\begin{equation}\begin{aligned}
\mathrm{d} s^{2}=&-\left(\frac{r-2 M}{r-2 D}\right) \mathrm{d} t^{2}+\left(\frac{r-2 M}{r-2 D}\right)^{-1} \mathrm{d} r^{2} \\
&+r(r-2 D)\left(\mathrm{d} \theta^{2}+\sin ^{2} \theta \mathrm{d} \phi^{2}\right).
\end{aligned}\end{equation}
Throughout this paper the natural units are set as $\hbar=G=c=k_B=1$. The massless Dirac equation can be written as $\gamma^{\alpha}e_{a}^{\mu}(\partial_\mu + \Gamma_\mu)\psi=0$, where $\gamma^{\alpha}$ is the Dirac matrix, $e_{a}^{\mu}$ corresponds to the inverse of the tetrad and $\Gamma_\mu$ is the spin connection coefficient. Solving massless Dirac equation near the event horizon leads to positive frequency outgoing solutions outside the region $I$ and inside regions $II$ as  \cite{Wang}
\begin{equation}\label{Dirac}
\psi_k^{\nu+}=\xi e^{\mp i \omega u},
\end{equation}
where $\nu \in \lbrace I, II \rbrace $ shows the regions, $k$ represents the field mode, $\xi$ is a 4-component Dirac spinor, and $\omega$ is a monochromatic frequency of the Dirac field. In Eq.(\ref{Dirac}), $u$ represents the retarded time and is defined as follows
\begin{equation}
u=t-2(M-D)\ln\left[ \frac{r-2M}{2M-2D}\right].
\end{equation}
\begin{figure}[t]
\center
  \includegraphics[width=0.45\textwidth]{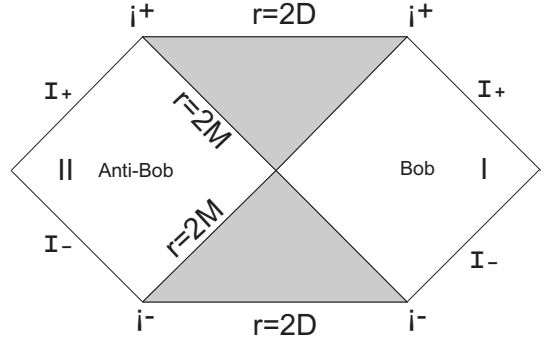}
\caption{The Penrose diagrams for the GHS dilation black hole which
shows the world-line of Bob and Anti-Bob. $i_0$ denotes the spatial
infinities, $i^{−}$ ($i^+$) represents time-like past (future) infinity. $I_{-}$ ($I_{+}$)
shows light-like past (future) infinity. }\label{kjk}
\end{figure}
To show the general form of the GHS dilation black hole, the Carter-Penrose diagrams for this space-times is plotted in Fig. \ref{kjk}. In diagram $r=2M$ shows the event horizons  and $r=2D$ represents the singularity of the black hole. $I$ and $II$ show the two general disconnected regions. The Dirac field can be quantized by using the complete orthogonal basis $\psi_k^{\nu+}$ as
\begin{equation}\label{GHS}
\psi_{\mathrm{out}}=\sum_{\nu=I, I I} \int \mathrm{d} k\left(a_{k}^{\nu} \psi_{k}^{\nu+}+b_{k}^{\nu *} \psi_{k}^{\nu-}\right)
\end{equation}
where $a_{k}^{\nu}$ and $b_{k}^{\nu *}$ are the fermion annihilation and anti-fermion creation operators respectively. 
One can use the generalized Kruskal coordinates to introduce the new orthogonal basis for positive energy mode as 
\begin{eqnarray}
\chi_{k}^{I+}&=&e^{2(M-D) \pi \omega} \psi_{k}^{I+}+e^{-2(M-D) \pi \omega} \psi_{-k}^{I I-}, \nonumber \\
\chi_{k}^{II+}&=&e^{-2(M-D) \pi \omega} \psi_{-k}^{I-}+e^{2(M-D) \pi \omega} \psi_{k}^{I I+}.
\end{eqnarray}
These new bases can be used to expand the Dirac fields in the Kruskal coordinates as
\begin{equation}\label{KRU}
\begin{aligned}
\psi_{\mathrm{out}}=& \sum_{\nu=I, I I} \int \mathrm{d} k | \frac{1}{\sqrt{2 \cosh [4(M-D) \pi \omega]}} \\
& \times\left(c_{k}^{\nu} \chi_{k}^{\nu+}+d_{k}^{\nu *} \chi_{k}^{\nu-}\right),
\end{aligned}\end{equation}
where $c_{k}^{\nu}$ and $d_{k}^{\nu*}$ are the fermion annihilation and antifermion creation operators which act on the Kruskal vacuum. Eq.(\ref{GHS}) is the expansion of the Dirac field in GHS dilation, while Eq.(\ref{KRU}) corresponds to the decomposition of the Dirac field in Kruskal modes. Using a suitable Bogoliubov transformation, one can obtain the annihilation operator $c_{k}^{\nu}$ as 
\begin{equation}c_{k}^{I}=\left(e^{-\frac{\omega}{T}}+1\right)^{-\frac{1}{2}} a_{k}^{I}-\left(e^{\frac{\omega}{T}}+1\right)^{-\frac{1}{2}} b_{k}^{I I *}
\end{equation}
where $T$ is the radiation temperature \cite{Huang,Martin,Bruschi}. The form of the ground state in the GHS dilation coordinates is regarded as a two-mode squeezed state in Kruskal coordinates. So, the vacuum and excited state can be expanded as follows
\begin{equation}
\begin{array}{l}
|0\rangle^{+}=x^{-\frac{1}{2}}|0\rangle_{I}^{+}|0\rangle_{I I}^{-}+y^{-\frac{1}{2}}|1\rangle_{I}^{+}|1\rangle_{I I}^{-} \\
|1\rangle^{+}=|1\rangle_{I}^{+}|0\rangle_{I I}^{-}
\end{array}
\end{equation}
where $x=\left(e^{-\frac{\omega}{T}}+1\right)$, $y=\left(e^{\frac{\omega}{T}}+1\right)$, $\vert n \rangle_{I}$ and $\vert n \rangle_{II}$ are the orthonormal basis of outside and inside the region of the event horizon respectively. The superscripts $+$ and $-$ show the particle and anti-particle respectively. In the following,   we abandon the superscripts $\pm$ in order to simplify the formulation.
Since the region $I$ and $II$ are completely disconnected, one can
obtain the physical accessible part $I$ by tracing over the state of the region $II$.  Interestingly, the whole process can be thought of as a quantum channel \cite{Huang}. In dynamics
of the open quantum system, the state of the system changes as a result of interaction with its surroundings. So, the loss of information in space-time with the event horizon can be considered as an open quantum system. Changing the state of an open quantum system at a given time is described by a dynamical map or quantum channel. The dynamical map $\Phi_t$ converts the initial state of the system $\rho_0 $ to the evolved state $\rho_t$ as $\rho_0 \rightarrow \rho_t=\Phi_t\rho_0$. The dynamical map can be written in Kraus form as $\rho_t=\Phi_t\rho_0=\sum_m W_m \rho_0 W_m^{\dag}$. If the initial state of the system is considered as $\rho_0=\sum_{i,j=0,1}\rho_{ij}\vert i\rangle\langle j \vert$, the dynamical map can be written as $\Phi_t \rho_0=\sum_{i,j=0,1}\varepsilon(\vert i\rangle\langle j \vert)\rho_{ij}$, where $\varepsilon(\vert i\rangle\langle j \vert)=\sum_m W_m (\vert i\rangle\langle j \vert)W_m^{\dag}$.  Considering the GHS dilation black hole, the dynamical map can be written as the Unruh channel,
\begin{equation}\label{channel}
\begin{aligned}
\varepsilon(|0\rangle\langle 0|) &=\left(e^{-\frac{\omega}{T}}+1\right)^{-1}|0\rangle_{I}\langle 0 |+\left(e^{\frac{\omega}{T}}+1\right)^{-1} | 1 \rangle_{I}\langle 1| \\
\varepsilon(|0\rangle\langle 1|) &=\left(e^{-\frac{\omega}{T}}+1\right)^{-\frac{1}{2}}|0\rangle_{I}\langle 1| \\
\varepsilon(|1\rangle\langle 0|) &=\left(e^{-\frac{\omega}{T}}+1\right)^{-\frac{1}{2}}|1\rangle_{I}\langle 0| \\
\varepsilon(|1\rangle\langle 1|) &=|1\rangle_{I}\langle 1|,
\end{aligned}\end{equation}
where the partial trace is done over the state of the interior region.
\begin{figure*}[h] 
     \begin{center}

    \hspace*{-0.25cm} \subfigure[]{%
            \label{figure2(a)}
            \includegraphics[height=6cm, width=9cm]{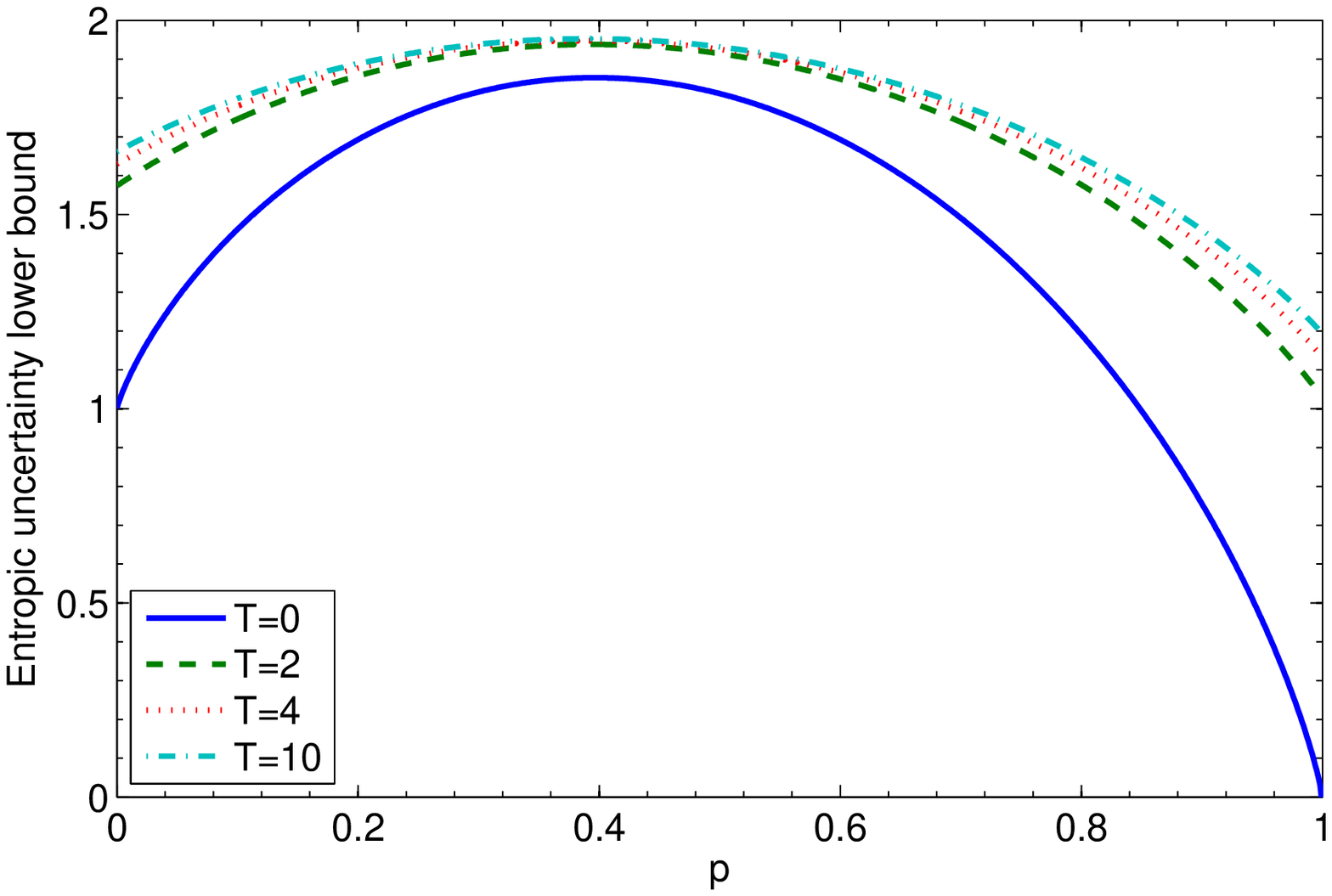}
        }%
        \hspace*{0.01cm} \subfigure[]{%
           \label{figure2(b)}
           \includegraphics[height=5.65cm, width=8.75cm]{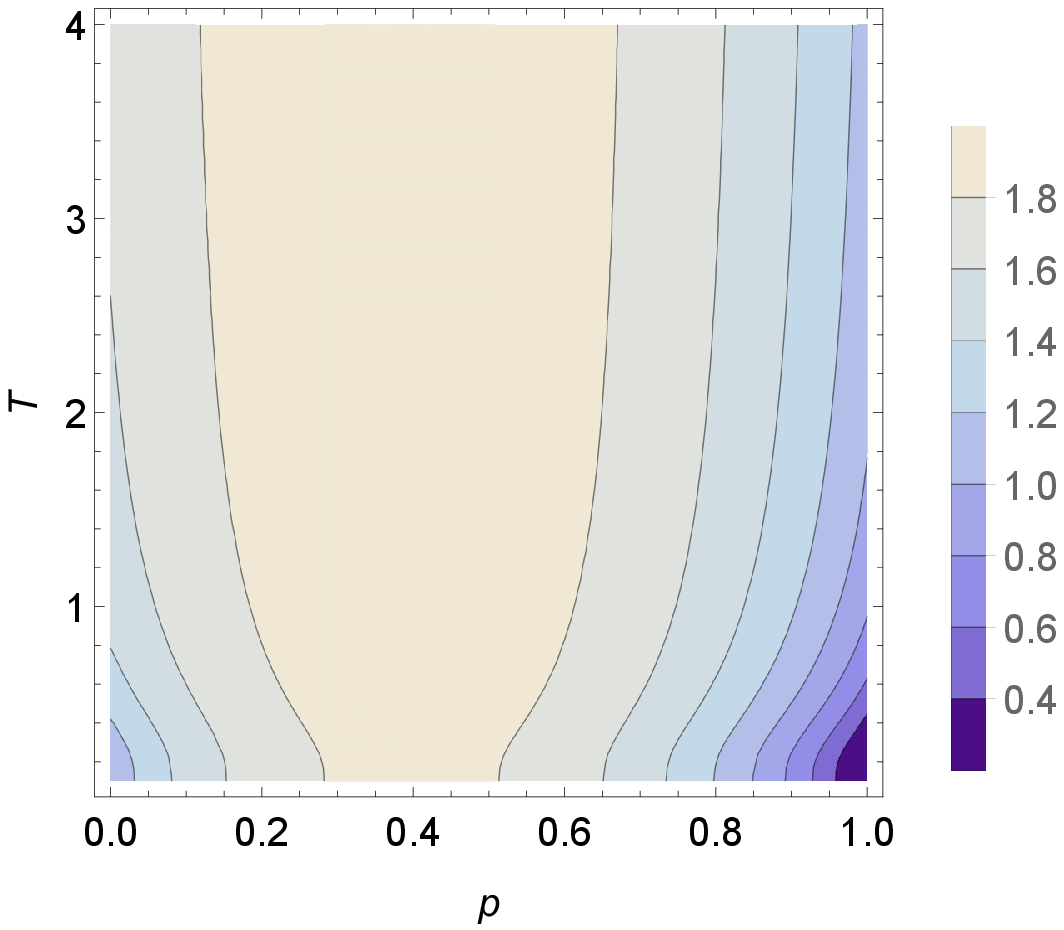}
        }\\ 

    \caption{%
        (a) Entropic uncertainty lower bound when Bob prepares a correlated bipartite state in a special class of state:$\rho^{AB}=p\left|\Psi^{-}\right\rangle \langle\Psi^{-} |+\frac{1-p}{2}\left(|\Psi^{+}\rangle\langle\Psi^{+}|+| \Phi^{+}\rangle \langle\Phi^{+}|\right)$ in terms of probability parameter $p$ for different values of Hawking temperature when $\omega=1$. (b) The contour plot of entropic uncertainty lower bound when Bob prepares a correlated bipartite state in a special class of state:$\rho^{AB}=p\left|\Psi^{-}\right\rangle \langle\Psi^{-} |+\frac{1-p}{2}\left(|\Psi^{+}\rangle\langle\Psi^{+}|+| \Phi^{+}\rangle \langle\Phi^{+}|\right)$ in terms of Hawking temperature $T$ and probability parameter $p$ when $\omega=1$.
     }%
     \end{center}
     \label{figure2}
\end{figure*}
In Fig.\ref{figure2(a)}, the entropic uncertainty lower bound is plotted in terms of probability parameter $p$ for different values of Hawking temperature. As can be seen, the entropic uncertainty lower bound increases  with increasing Hawking temperature. This is what we expected, the quantum correlation decreases under the influence of Hawking radiation and so the entropic uncertainty increases with increasing Hawking temperature \cite{80,81,82,83,84}. Fig.\ref{figure2(b)} shows the contour plot of entropic uncertainty lower bound in terms of Hawking temperature $T$ and probability parameter $p$. As can be seen, from Fig.\ref{figure2(b)}, at different Hawking temperatures the entropic uncertainty lower bound has its lowest value for the case in which $p=1$ and the state is maximally entangled. It is also observed that for different values of $p$ this entropic uncertainty lower bound increases with increasing Hawking temperature.
\begin{figure*}[h] 
     \begin{center}

    \hspace*{-0.25cm} \subfigure[]{%
            \label{figure3(a)}
            \includegraphics[height=6cm, width=9cm]{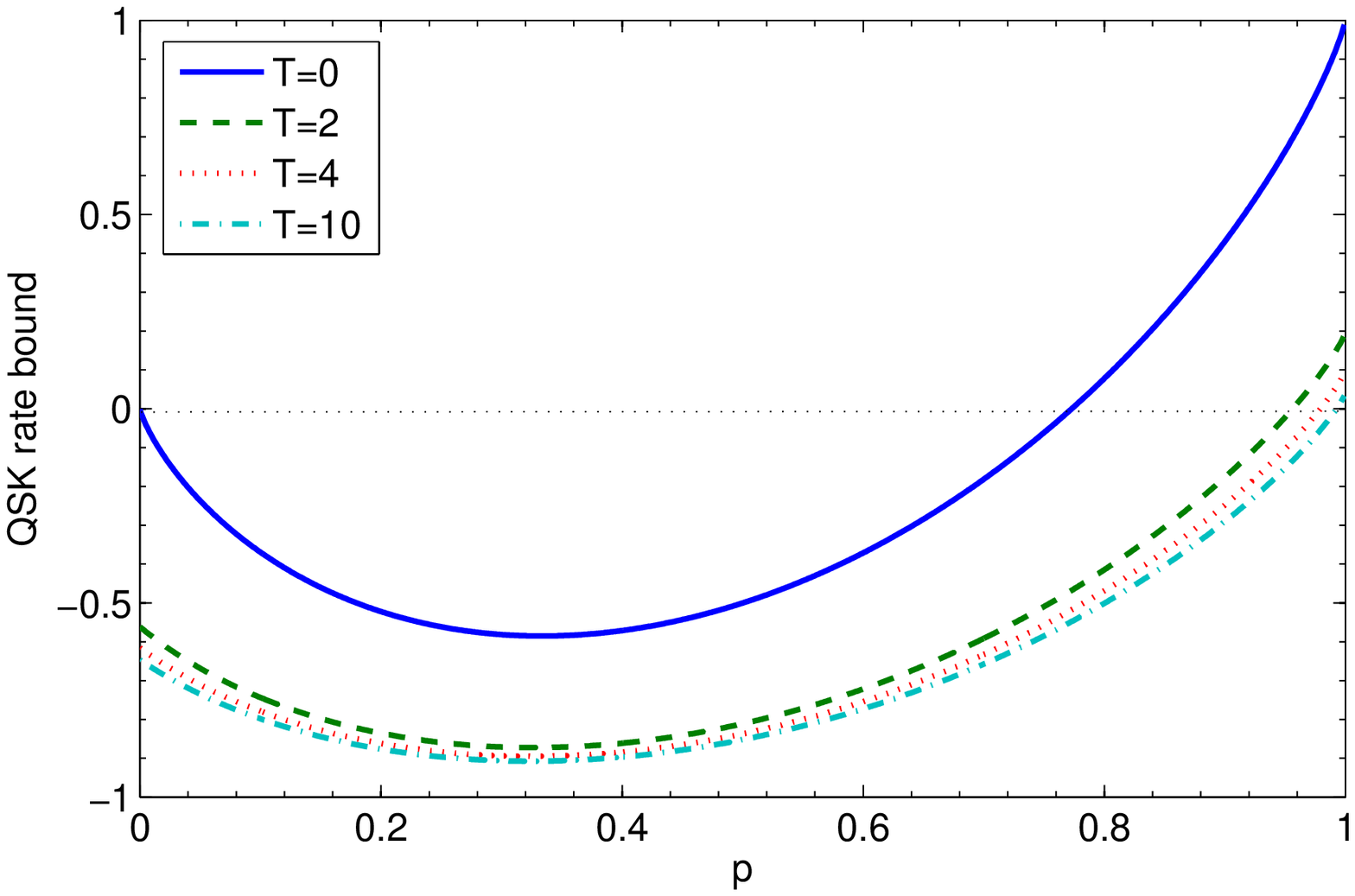}
        }%
        \hspace*{0.01cm} \subfigure[]{%
           \label{figure3(b)}
           \includegraphics[height=5.65cm, width=8.75cm]{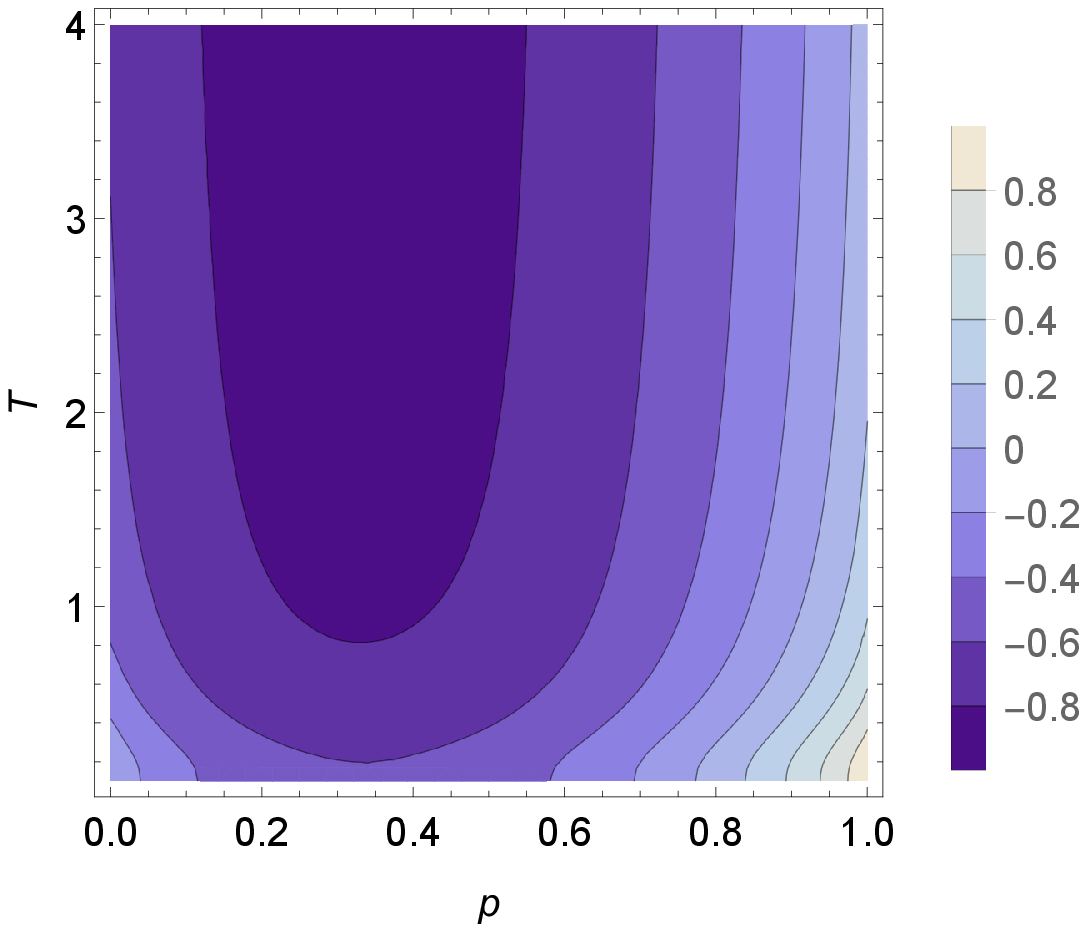}
        }\\ 

    \caption{%
        (a) QSK rate bound when Bob prepares a correlated bipartite state in a special class of state:$\rho^{AB}=p\left|\Psi^{-}\right\rangle \langle\Psi^{-} |+\frac{1-p}{2}\left(|\Psi^{+}\rangle\langle\Psi^{+}|+| \Phi^{+}\rangle \langle\Phi^{+}|\right)$ in terms of probability parameter $p$ for different values of Hawking temperature when $\omega=1$. (b) The contour plot of QSK rate bound when Bob prepares a correlated bipartite state in a special class of state:$\rho^{AB}=p\left|\Psi^{-}\right\rangle \langle\Psi^{-} |+\frac{1-p}{2}\left(|\Psi^{+}\rangle\langle\Psi^{+}|+| \Phi^{+}\rangle \langle\Phi^{+}|\right)$ in terms of Hawking temperature $T$ and probability parameter $p$ when $\omega=1$.
     }%
     \end{center}
     \label{figure3}
\end{figure*}
\section{Entropic uncertainty lower bound in Garfinkle-Horowitz-Strominger dilation black hole}\label{sec3}
In this section the uncertainty game between Alice and Bob in Garfinkle-Horowitz-Strominger dilation black hole is investigated.  At the beginning of the game, Bob prepares a correlated bipartite state $\rho_{AB}$ then sends the first part to Alice and keeps the other part as a quantum memory memory. In this step of the game, both of them free falling towards the black hole. In the next step, Alice remains free
falling into the black hole while Bob is in a fixed position outside the black hole. Then Alice measures one of the two observable $Q$ or $R$ on her state and sends her  measurement choice to Bob via a classical
communication channel. The main purpose of this game for Bob is to reduce his uncertainty about the result of Alice's measurement. If Bob can correctly guess the result of Alice's measurement in this situation, he will win this game. Due to the fact that the resident observer cannot access modes beyond the event horizon, the lost information reduces the entanglement between Alice and Bob. So, it changes the uncertainty bound. By reducing Bob's distance from the black hole's event horizon, his uncertainty about the result of Alice's measurement will increase.
\subsection{Examples}
\subsubsection{Bell diagonal state}
As a first example, let us consider the case in which Alice and Bob share the set of two-qubit states with the maximally mixed marginal  states. This state is defined as follows

\begin{equation}\label{Bel}
\rho^{A B}=\frac{1}{4}\left(I \otimes I+\sum_{i=1}^{3} w_{i j} \sigma_{i} \otimes \sigma_{j}\right)\end{equation}

where $\sigma_i$($i=1,2,3$) are the Pauli matrices. Using the singular value decomposition, the matrix $W=\lbrace w_ij \rbrace$ can be diagonalized by a local unitary transformation. So, Eq. (\ref{Bel}) can be rewritten as 
\begin{equation}
\rho^{AB}=\frac{1}{4}(I \otimes I + \sum_i r_i \sigma_i \otimes \sigma_i),
\end{equation} 
where $\vec{r}=(r_1,r_2,r_3)$ is limited to  a tetrahedron defined by the set of vertices ($-1,-1,-1$),
  ($-1,1,1$),  ($1,-1,1$) and ($1,1,-1$). We consider the case in which $r_1=1-2p$, $r_2=-p$ and $r_3=-p$ where $0 \leq p \leq 1$. So the state in Eq.(\ref{Bel1}) can be written as 
\begin{equation}\label{Bel1}
\rho^{AB}=p\left|\Psi^{-}\right\rangle \langle\Psi^{-} |+\frac{1-p}{2}\left(|\Psi^{+}\rangle\langle\Psi^{+}|+| \Phi^{+}\rangle \langle\Phi^{+}|\right),
\end{equation}
where $\vert \Psi^{\pm} \rangle = \frac{1}{\sqrt{2}}(\vert 01 \rangle \pm \vert 10 \rangle)$ and $\vert \Phi^{\pm} \rangle = \frac{1}{\sqrt{2}}(\vert 00 \rangle \pm \vert 11 \rangle)$. We consider a model in which the memory particle $B$ (At the disposal of Bob) is located near the event horizon outside the black hole, while the measured particle $A$ (At the disposal of Alice) is free falling.  The effect of Hawking radiation can be defined by applying the Unruh channel in Eq.(\ref{channel}) on Bob's state.

\begin{figure*}[h] 
     \begin{center}

    \hspace*{-0.25cm} \subfigure[]{%
            \label{figure4(a)}
            \includegraphics[height=6cm, width=9cm]{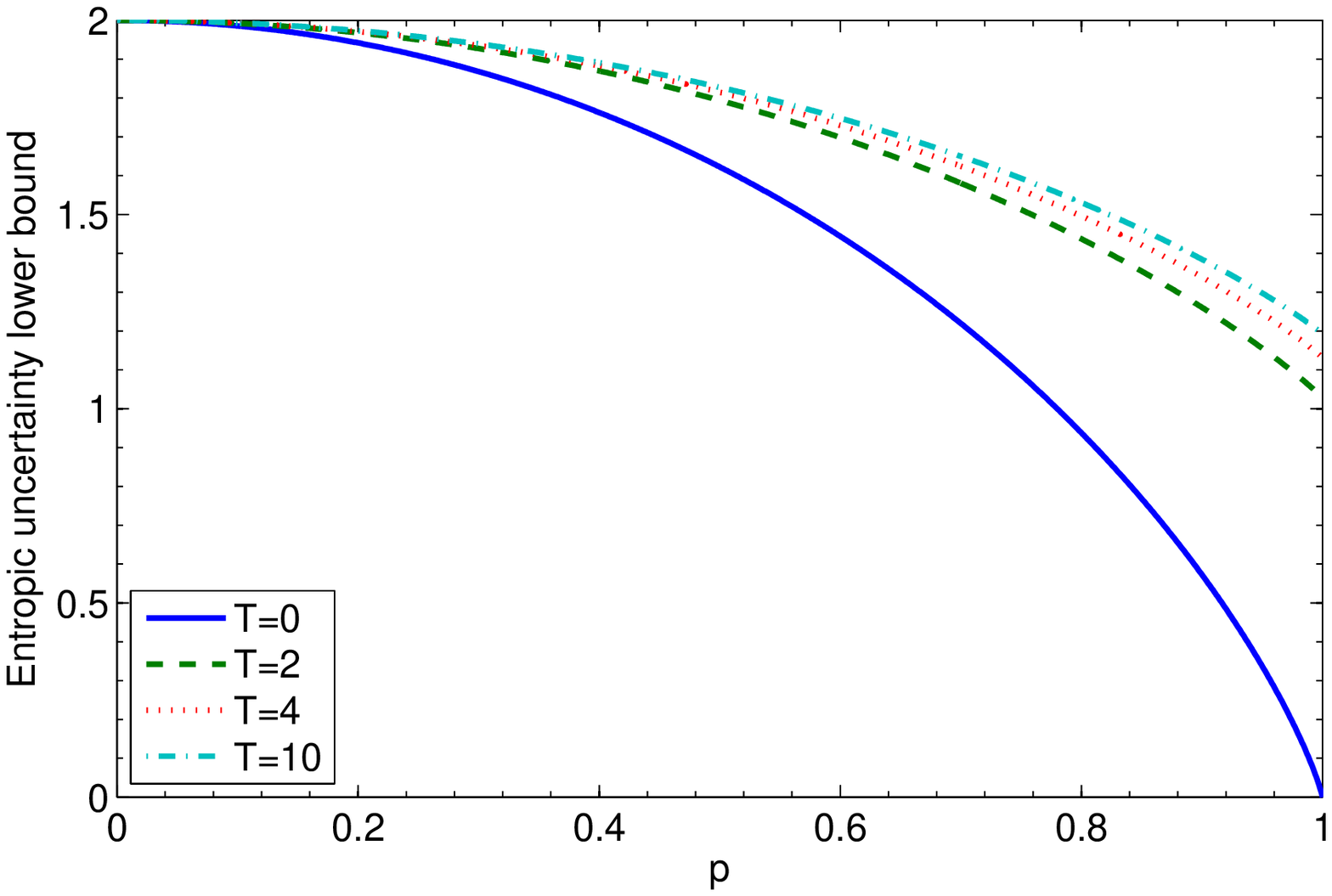}
        }%
        \hspace*{0.01cm} \subfigure[]{%
           \label{figure4(b)}
           \includegraphics[height=5.65cm, width=8.75cm]{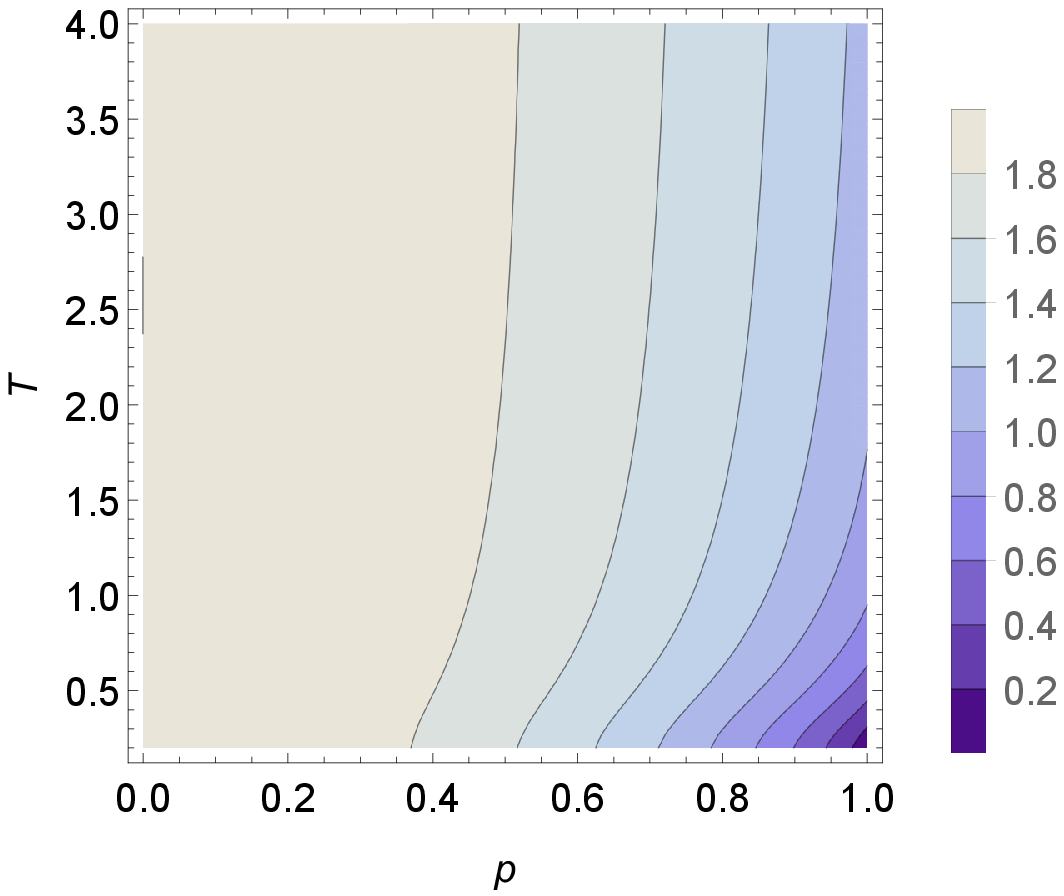}
        }\\ 

    \caption{%
        (a) Entropic uncertainty lower bound when Bob prepares a correlated bipartite state in a special class of state:$\rho^{AB}=\frac{1-p}{4}I\otimes I+p \vert \psi^{-} \rangle \langle \Psi^{-} \vert$ in terms of probability parameter $p$ for different values of Hawking temperature when $\omega=1$. (b) The contour plot of entropic uncertainty lower bound when Bob prepares a correlated bipartite state in a special class of state:$\rho^{AB}=\frac{1-p}{4}I\otimes I+p \vert \psi^{-} \rangle \langle \Psi^{-} \vert$ in terms of Hawking temperature $T$ and probability parameter $p$ when $\omega=1$.
     }%
     \end{center}
     \label{figure4}
\end{figure*}
\begin{figure*}[h] 
     \begin{center}

    \hspace*{-0.25cm} \subfigure[]{%
            \label{figure5(a)}
            \includegraphics[height=6cm, width=9cm]{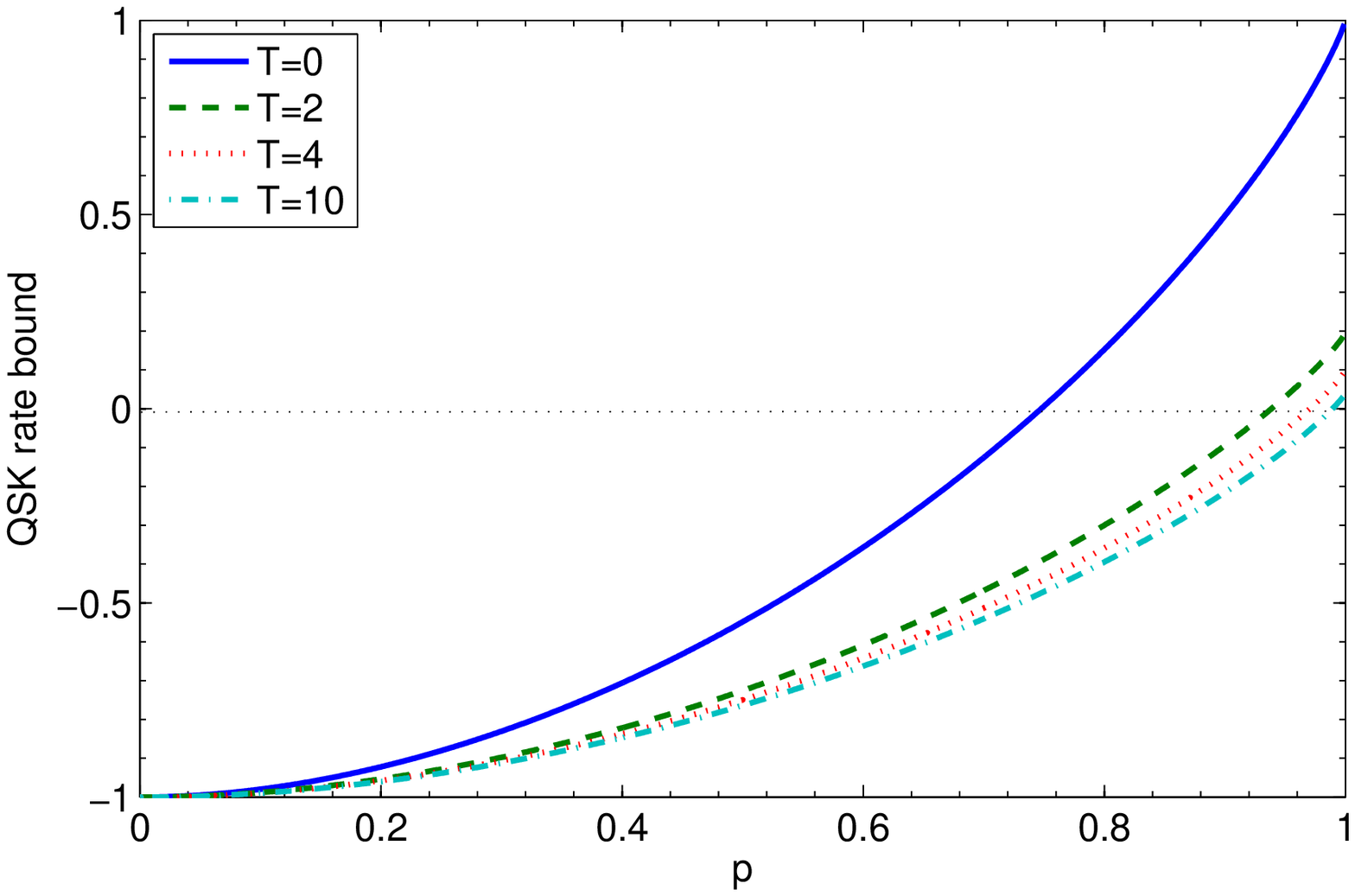}
        }%
        \hspace*{0.01cm} \subfigure[]{%
           \label{figure5(b)}
           \includegraphics[height=5.65cm, width=8.75cm]{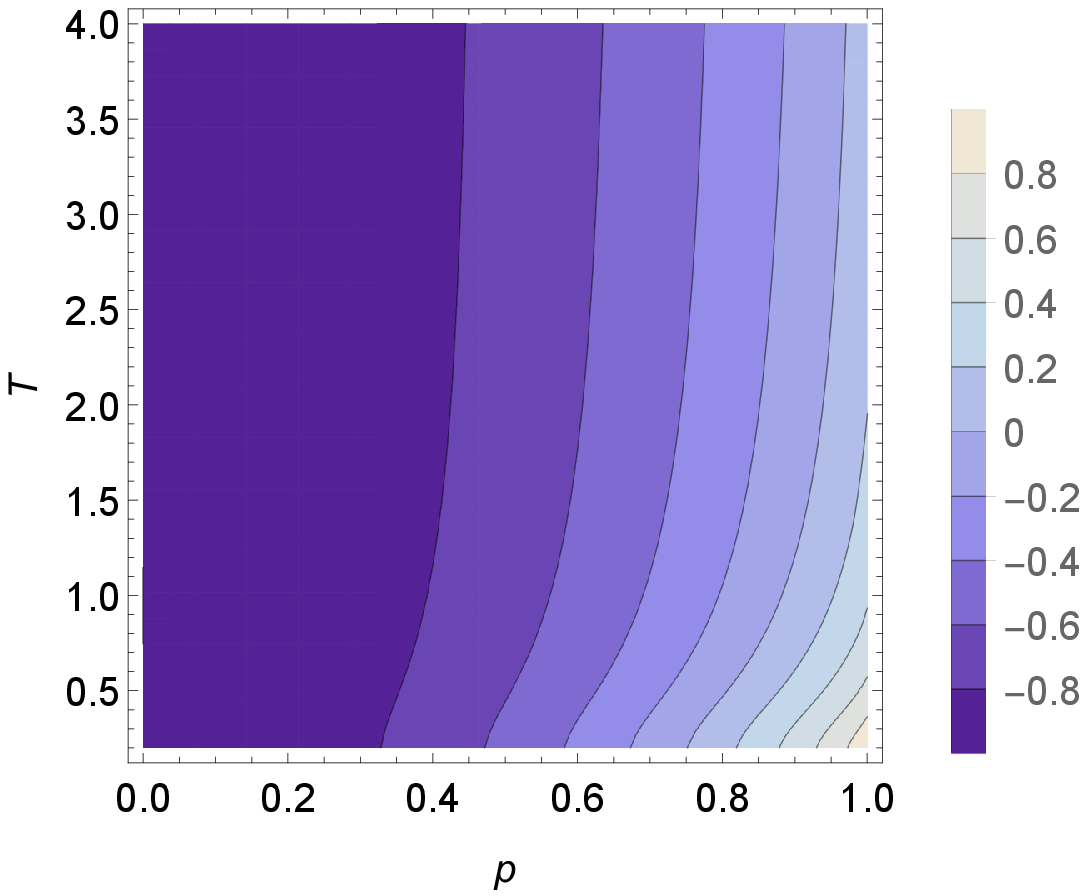}
        }\\ 

    \caption{%
        (a) QSK rate bound when Bob prepares a correlated bipartite state in a special class of state:$\rho^{AB}=\frac{1-p}{4}I\otimes I+p \vert \psi^{-} \rangle \langle \Psi^{-} \vert$ in terms of probability parameter $p$ for different values of Hawking temperature when $\omega=1$. (b) The contour plot of QSK rate bound when Bob prepares a correlated bipartite state in a special class of state:$\rho^{AB}=\frac{1-p}{4}I\otimes I+p \vert \psi^{-} \rangle \langle \Psi^{-} \vert$ in terms of Hawking temperature $T$ and probability parameter $p$ when $\omega=1$.
     }%
     \end{center}
     \label{figure5}
\end{figure*}
\begin{figure*}[t] 
     \begin{center}

    \hspace*{-0.25cm} \subfigure[]{%
            \label{figure6(a)}
            \includegraphics[height=6cm, width=9cm]{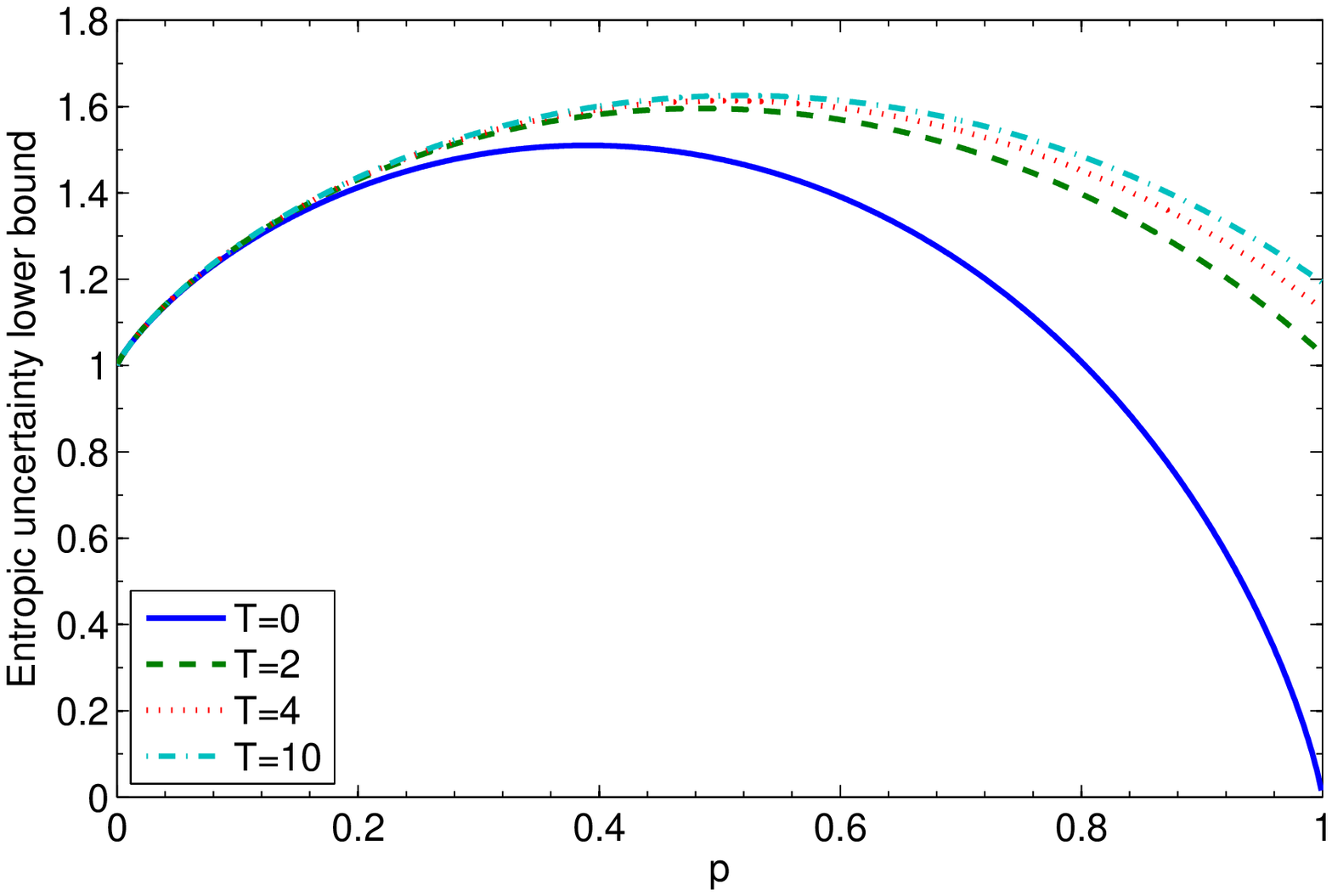}
        }%
        \hspace*{0.01cm} \subfigure[]{%
           \label{figure6(b)}
           \includegraphics[height=5.65cm, width=8.75cm]{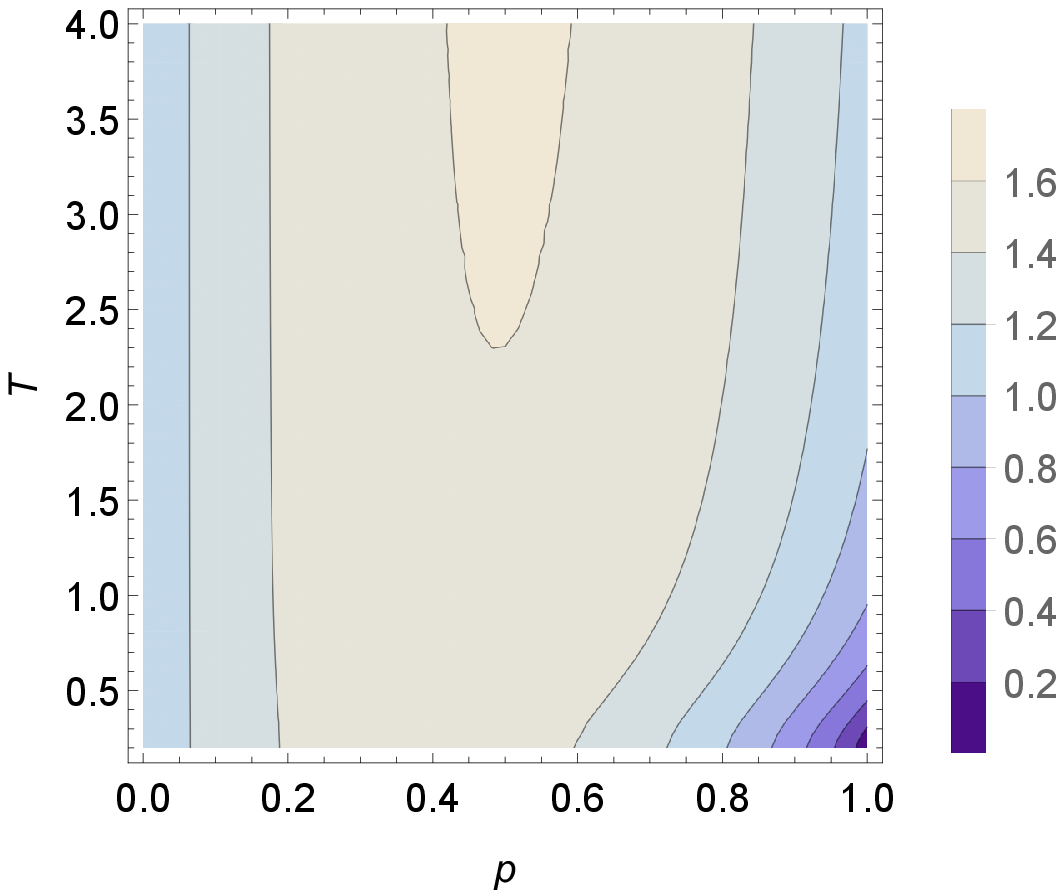}
        }\\ 

    \caption{%
        (a) Entropic uncertainty lower bound when Bob prepares a correlated bipartite state in a special class of state:$\rho^{AB}=p |\Psi^{+} \rangle \langle \Psi^{+}| + (1-p)| 11 \rangle \langle 11$ in terms of probability parameter $p$ for different values of Hawking temperature when $\omega=1$. (b) The contour plot of entropic uncertainty lower bound when Bob prepares a correlated bipartite state in a special class of state:$\rho^{AB}=p |\Psi^{+} \rangle \langle \Psi^{+}| + (1-p)| 11 \rangle \langle 11$ in terms of Hawking temperature $T$ and probability parameter $p$ when $\omega=1$.
     }%
     \end{center}
     \label{figure6}
\end{figure*}
\begin{figure*}[t] 
     \begin{center}

    \hspace*{-0.25cm} \subfigure[]{%
            \label{figure7(a)}
            \includegraphics[height=6cm, width=9cm]{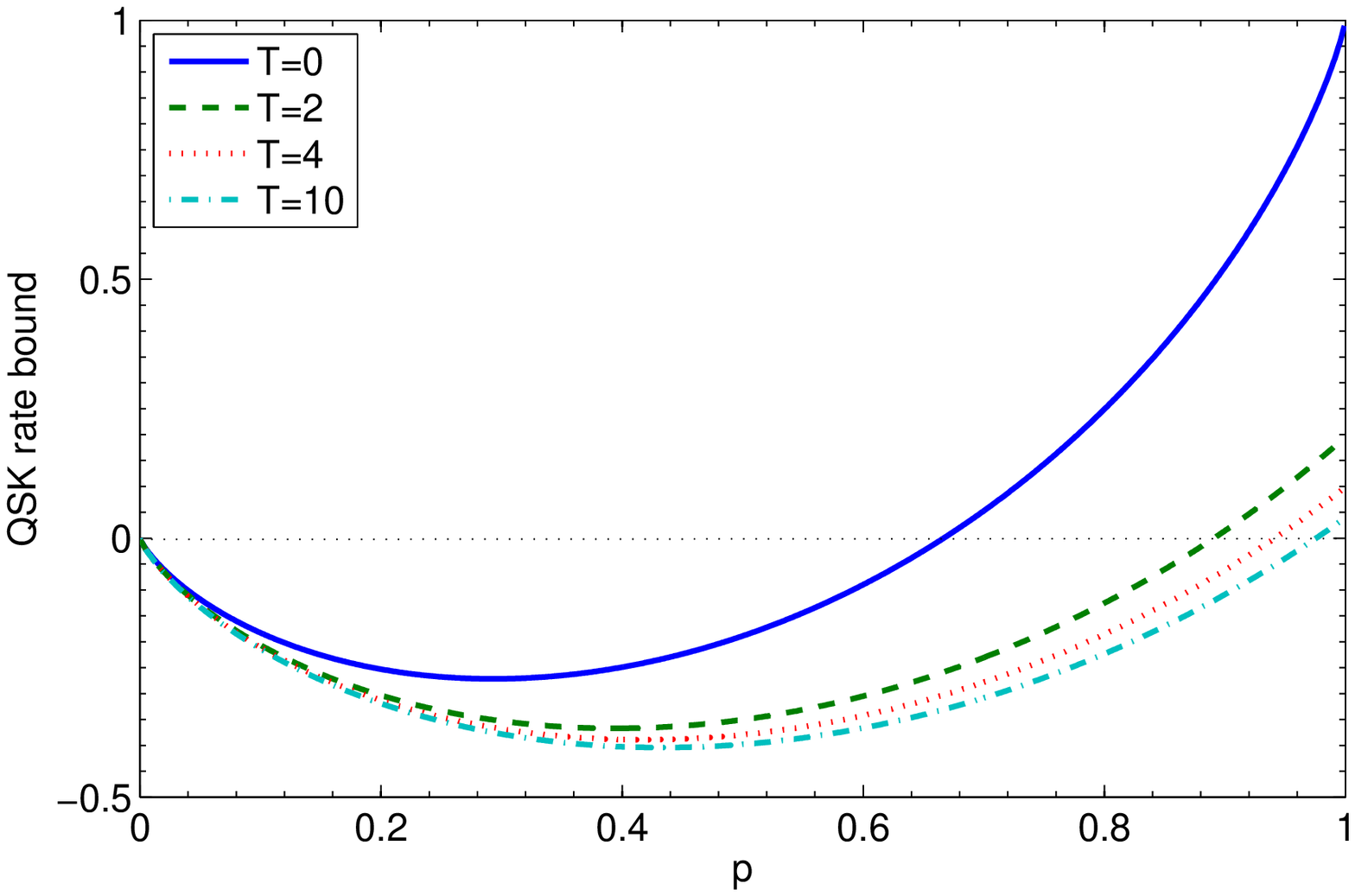}
        }%
        \hspace*{0.01cm} \subfigure[]{%
           \label{figure7(b)}
           \includegraphics[height=5.65cm, width=8.75cm]{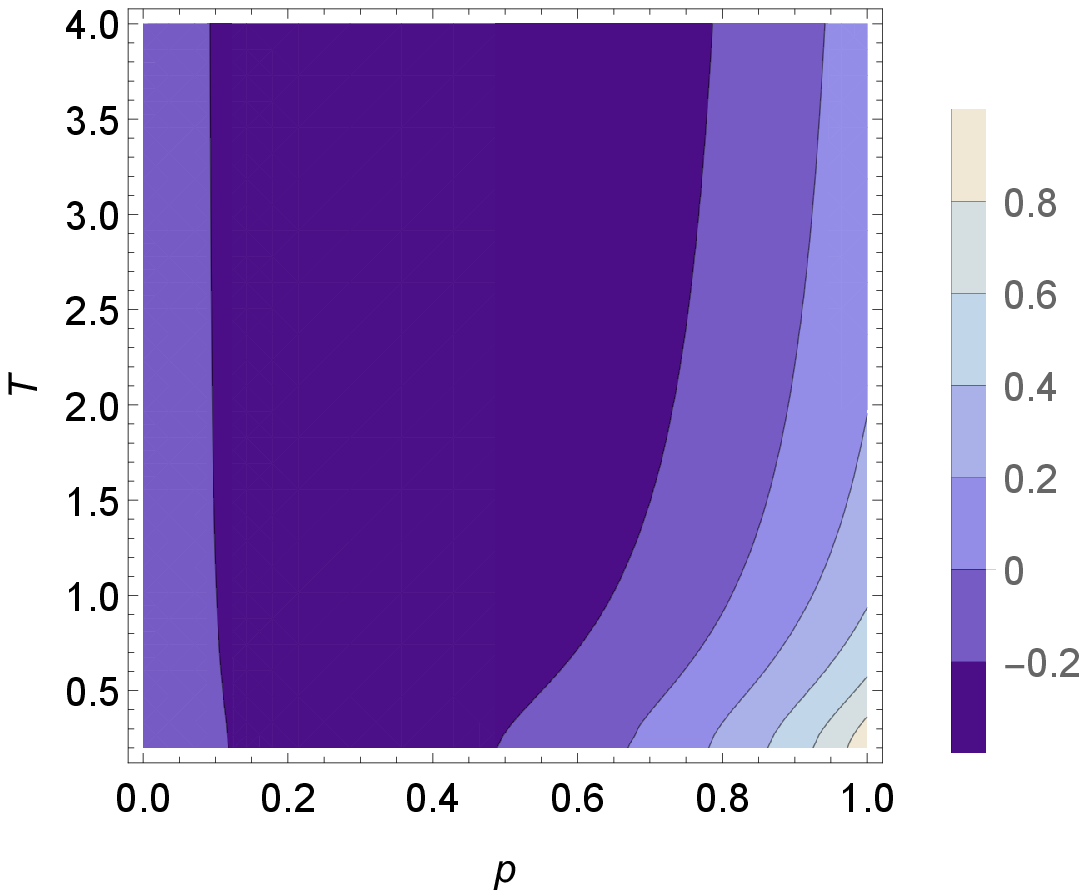}
        }\\ 

    \caption{%
        (a) QSK rate bound when Bob prepares a correlated bipartite state in a special class of state:$\rho^{AB}=\frac{1-p}{4}I\otimes I+p \vert \psi^{-} \rangle \langle \Psi^{-} \vert$ in terms of probability parameter $p$ for different values of Hawking temperature when $\omega=1$. (b) The contour plot of QSK rate bound when Bob prepares a correlated bipartite state in a special class of state:$\rho^{AB}=\frac{1-p}{4}I\otimes I+p \vert \psi^{-} \rangle \langle \Psi^{-} \vert$ in terms of Hawking temperature $T$ and probability parameter $p$ when $\omega=1$.
     }%
     \end{center}
     \label{figure7}
\end{figure*}
In Fig.\ref{figure3(a)}, the QSK rate bound is plotted in terms of probability parameter $p$ for different values of Hawking temperature. As can be seen, the QSK rate bound decreases  with increasing Hawking temperature. This is what we expected, the quantum correlation decreases under the influence of Hawking radiation and so the QSK rate bound decreases with increasing Hawking temperature. Fig.\ref{figure3(b)} shows the contour plot of QSK rate bound in terms of Hawking temperature $T$ and probability parameter $p$. As can be seen, from Fig.\ref{figure3(b)}, at different Hawking temperatures the QSK rate bound has its highest value for the case in which $p=1$ and the state is maximally entangled. It is also observed that for different values of $p$ the QSK rate bound decreases with increasing Hawking temperature. As can be seen, from Figs.\ref{figure3(a)} and \ref{figure3(b)}, the QSK rate bound is negative for some values of $p$ and T. So, one can conclude that for these values of $p$ and $T$ the states are not
good enough to support quantum key distribution protocols.
\subsubsection{Werner state}
As a second example, Let us consider the case in which Alice and Bob initially share a two-qubit Werner state 
\begin{equation}
\rho^{AB}=\frac{1-p}{4}I\otimes I+p \vert \psi^{-} \rangle \langle \Psi^{-} \vert,
\end{equation}
where $0\leq p \leq 1$.

In Fig.\ref{figure4(a)}, the entropic uncertainty lower bound is plotted in terms of probability parameter $p$ for different values of Hawking temperature. As can be seen, the entropic uncertainty lower bound increases  with increasing Hawking temperature. Fig.\ref{figure4(b)} shows the contour plot of entropic uncertainty lower bound in terms of Hawking temperature $T$ and probability parameter $p$. As can be seen, from Fig.\ref{figure4(b)}, at different Hawking temperatures the entropic uncertainty lower bound has its lowest value for the case in which $p=1$ and the state is maximally entangled. It is observed that for different values of $p$ this entropic uncertainty lower bound increases with increasing Hawking temperature. It is also observed that for the case in which $p=0$, the entropic uncertainty lower bound not affected by Hawking radiation.

In Fig.\ref{figure5(a)}, the QSK rate bound is plotted in terms of probability parameter $p$ for different values of Hawking temperature. As can be seen, the QSK rate bound decreases  with increasing Hawking temperature. This is what we expected, the quantum correlation decreases under the influence of Hawking radiation and so the QSK rate bound decreases with increasing Hawking temperature. Fig.\ref{figure5(b)} shows the contour plot of QSK rate bound in terms of Hawking temperature $T$ and probability parameter $p$. As can be seen, from Fig.\ref{figure5(b)}, at different Hawking temperatures the QSK rate bound has its highest value for the case in which $p=1$ and the state is maximally entangled. It is also observed that for different values of $p$ the QSK rate bound decreases with increasing Hawking temperature. As can be seen, from Figs.\ref{figure5(a)} and \ref{figure5(b)}, the QSK rate bounds is negative for some values of $p$ and T. So, one can conclude that for these values of $p$ and $T$ the states are not
good enough to support quantum key distribution protocols. It is also observed that for the case in which $p=0$, the QSK rate bound not affected by Hawking radiation.

\subsubsection{Two-qubit X states}
As the last example, let us consider the case in which Alice and Bob share a special class of two qubit X states
\begin{equation}
\rho^{AB}=p |\Psi^{+} \rangle \langle \Psi^{+}| + (1-p)| 11 \rangle \langle 11 ,
\end{equation}
where $0\leq p \leq 1$.

In Fig.\ref{figure6(a)}, the entropic uncertainty lower bound is plotted in terms of probability parameter $p$ for different values of Hawking temperature. The entropic uncertainty lower bound increases  with increasing Hawking temperature. Fig.\ref{figure6(b)} shows the contour plot of entropic uncertainty lower bound in terms of Hawking temperature $T$ and probability parameter $p$. As can be seen, from Fig.\ref{figure6(b)}, at different Hawking temperatures the entropic uncertainty lower bound has its lowest value for the case in which $p=1$ and the state is maximally entangled. It is observed that for different values of $p$ this entropic uncertainty lower bound increases with increasing Hawking temperature. It is also observed that for the case in which $p=0$, the entropic uncertainty lower bound not affected by Hawking radiation.

In Fig.\ref{figure7(a)}, the QSK rate bound is plotted in terms of probability parameter $p$ for different values of Hawking temperature. As can be seen, the QSK rate bound decreases  with increasing Hawking temperature. This is what we expected, the quantum correlation decreases under the influence of Hawking radiation and so the QSK rate bound decreases with increasing Hawking temperature. Fig.\ref{figure7(b)} shows the contour plot of QSK rate bound in terms of Hawking temperature $T$ and probability parameter $p$. As can be seen, from Fig.\ref{figure7(b)}, at different Hawking temperatures the QSK rate bound has its highest value for the case in which $p=1$ and the state is maximally entangled. It is also observed that for different values of $p$ the QSK rate bound decreases with increasing Hawking temperature. As can be seen, from Figs.\ref{figure7(a)} and \ref{figure7(b)}, the QSK rate bound is negative for some values of $p$ and T. So, one can conclude that for these values of $p$ and $T$ the states are not
good enough to support quantum key distribution protocols. It is also observed that for the case in which $p=0$, the QSK rate bound not affected by Hawking radiation.
\section{Conclusion}\label{conclusion}\label{sec4}
In this work we studied the entropic uncertainty relation in Garfinkle-Horowitz-Strominger dilation black hole. For this purpose, we consider the uncertainty game between Alice and Bob. At first Bob prepares a correlated bipartite state $\rho_{AB}$ then he sends the first part $A$ to Alice and keeps the other part as a quantum memory memory $B$. In this step of the game, both of them free falling towards the black hole. In the next step, Alice remains free falling into the black hole while Bob is in a fixed position outside the black hole. Then Alice measures one of the two observable $Q$ or $R$ on her state and sends her  measurement choice to Bob via a classical communication channel. The main purpose of this game for Bob is to reduce his uncertainty about the result of Alice's measurement. As mentioned before, quantum correlations are reduced by the effect of Hawking radiation. Therefore, due to the inverse relation between quantum correlation and uncertainty the uncertainty bound will increase as a result of Hawking effect. We also investigated the effects of the Hawking radiation on QSK rate bound. It was shown that the QSK rate bound decreases by increasing Hawking temperature.

\end{document}